\documentclass[12pt,3p]{elsarticle}

\makeatletter
\@ifclasswith{elsarticle}{review}{
    \usepackage{etoolbox}
    
    \AtBeginDocument{\linespread{1.5}\selectfont}
    
    \AtBeginEnvironment{abstract}{\linespread{1.5}\selectfont}
    
    \apptocmd{\maketitle}{\linespread{1.5}\selectfont}{}{}
}{
}
\makeatother

\biboptions{sort&compress}
\makeatletter
\def\ps@pprintTitle{%
  \let\@oddhead\@empty
  \let\@evenhead\@empty
  \def\@oddfoot{\reset@font\hfil\thepage\hfil}
  \let\@evenfoot\@oddfoot
}
\makeatother
\usepackage[utf8]{inputenc}
\usepackage[T2A,T1]{fontenc}
\usepackage[english]{babel}

\usepackage{amsmath,amssymb}
\usepackage{bm}
\usepackage {graphicx}
\graphicspath{{fig/}}
\usepackage{hyperref}
\usepackage{array}
\allowdisplaybreaks[4]

\usepackage[draft]{changes}
\definechangesauthor[color=violet]{ap}
\definechangesauthor[color=olive]{mb}
\begin{document}

\title{Multipolar exchange in a many-body homonuclear mixture of atoms in different internal states}
\author[1]{M.~Bulakhov}
\ead{bulakh@kipt.kharkov.ua}
\author[1]{A.S.~Peletminskii}
\ead{aspelet@kipt.kharkov.ua}
\author[1,2,3]{Yu.V.~Slyusarenko}
\affiliation[1]{
orgsnization={Akhiezer Institute for Theoretical Physics, National Science Center "Kharkiv Institute of Physics and Technology", NAS of Ukraine}, 
city={Kharkiv}, 
postcode={61108},  
country={Ukraine}}
\affiliation[2]{
organization={V.N. Karazin Kharkiv National University},
city={Kharkiv},
postcode={61022},
country={Ukraine}
}
\affiliation[3]{
    organization={Lviv Polytechnic National University},
    city={Lviv},
    postcode={79000}, 
    country={Ukraine}
}

\begin{abstract}
We develop a general method for constructing the many-body Hamiltonian of pairwise interactions describing homonuclear mixtures of atoms occupying states with different total angular momenta or other quantum numbers.
This Hamiltonian explicitly accounts for isotropic interactions governed by low-energy $s$-wave scattering. 
The advantage of the irreducible spherical tensor operator formalism is demonstrated: these operators give the Hamiltonian an explicit physical structure, account for all scattering channels, and include multipolar exchange interactions. The latter correspond to the exchange of both angular-momentum projections and the total angular momentum. 
Particular realizations of the general Hamiltonian, widely used in the physics of ultracold gases, are also analyzed. The resulting Hamiltonian provides a universal framework for investigating a broad range of quantum many-body phenomena in bosonic and fermionic atomic gases.

\end{abstract}
\maketitle
\section{Introduction}
    
Since the advent of optical traps, experiments with ultracold atomic gases have made significant progress in studying relatively stable mixtures of interacting homonuclear atoms \cite{Dalibard_PRL_2025,Baroni_NaturePhysRev_2024,Xue_PRA_2022}.
Such traps enable controlled population of different internal atomic states and allow efficient investigation of the properties of these mixtures.
Until recently, experiments have mainly focused on interaction effects between atoms in states with the same total angular momentum but different projections.
These systems have also been extensively studied theoretically in the context of spinor condensates \cite{Ohmi_JPSJ_1998,Ho_PRL_1998,Bigelow_PRL_1998,Akhiezer_JETP_1998,Ueda_PRA_2007,Peletminskii_PhysA_2007,Ueda_PhysRep_2012}, Cooper pairing \cite{Ho_PRL_1999,Vanderbos_PRX_2018}, collective excitations \cite{Yip_PRA_1999,Lewenstein_PRL_2013,Bulakhov_JPhysA_2023,Bulakhov_AnnPhys_2025}, and novel superfluid states in optical lattices \cite{Lecheminant_PRL_2005,Rapp_PRL_2007}.
More recently, increasing attention has been devoted to homonuclear gaseous mixtures in optical traps where atoms occupy internal states characterized not only by different values of the total angular momentum but also by other quantum numbers \cite{Baroni_NaturePhysRev_2024,Cominotti_EPL_2024}.
To the best of our knowledge, a consistent theoretical description of interatomic interactions in such systems remains incomplete.
In this work, we address this problem by constructing a many-body Hamiltonian that incorporates both direct and all possible exchange interaction processes between atoms in different internal states, in particular, with different total angular momenta.

To investigate these interaction processes in ultracold dilute gases under weak external magnetic fields, it is sufficient to describe the atomic states in the coupled basis, i.e., in terms of the total angular momentum and its projection.
Once the atomic states have been specified, the interaction Hamiltonian can be constructed using the formalism of irreducible spherical tensor operators (see, e.g., the classical monographs \cite{Wigner_1959,Racah_1959,Brink_1968,Edmonds_1996,Sakurai_2020}).
This approach has proven highly effective in such areas as nuclear magnetic resonance physics \cite{vanBeek_JChemPhys_2005}, quantum chemistry \cite{Krems_JPhysChemA_2004}, and atomic spectroscopy \cite{Casini_ApJ_2025}.
The use of spherical tensor operators endows the Hamiltonian with a transparent physical structure and allows one to account for all scattering channels and multipolar interactions.

It should be noted that within the framework of quantum chemistry and Feshbach resonance studies, homonuclear atomic mixtures also play a central role \cite{Krems_book_2009,Grimm_RevModPhys_2010}.
However, this area of study represents the case, which is opposite to our one, since the interactions are governed by strong molecular forces, intrinsic atomic and external magnetic fields whose energies exceed the hyperfine splitting. 
Under these conditions, the total angular momentum is no longer a good quantum number. 
Consequently, an uncoupled basis is often employed to describe the atoms, treating the orbital angular momentum and spin independently \cite{Krems_JPhysChemA_2004,Krems_InRevPhysChem_2005}.

In our view, homonuclear mixtures of atoms prepared in long-lived metastable states possess significant experimental potential for implementing the approach proposed here.
Notable examples include: titanium ($\approx 28\cdot 10^3$ s) \cite{Eustice_PhysRev_2025}, helium ($\approx 8000$ s) \cite{Hodgman_PRL_2009}, strontium ($\approx 520$ s) \cite{Yasuda_PhysRevLett_2004}, and ytterbium ($\approx 2$ s) \cite{Ishiyama_PhysRevLett_2023}.
Another relevant example is provided by gases of atoms with large magnetic dipole moments, such as erbium and dysprosium \cite{Lu_PhysRevLett_2010,Lepers_PhysRevLett_2018,Baier_PhysRevLett_2018}, which are characterized by a rich multipolar structure.
Certainly, a well-studied, albeit simpler, instance is provided by mixtures of hyperfine sublevels \cite{Cominotti_EPL_2024}.
Finally, Rydberg atoms also possess large dipole moments; however, the use of such mixtures is less promising due to their extremely short lifetimes.

The paper is organized as follows.
In Sec. \ref{sec:GeneralHamiltonian}, we construct the many-body Hamiltonian of pairwise interactions for homonuclear mixtures of atoms occupying states with different total angular momenta and other quantum numbers.
In Subsec. \ref{ssec:Aspects}, we discuss the description of internal atomic states and the multipolar representation of their interactions.
Subsec. \ref{ssec:STO} provides the justification for employing the formalism of irreducible spherical tensor operators.
In Subsec. \ref{subsec:InterHam}, we present a systematic construction of the interaction Hamiltonian and analyze exchange processes involving the transfer of both the projections and the total angular momentum of the atoms.
In Secs. \ref{sec:SpecificEqualMomenta} and \ref{sec:SpecificMixture}, we illustrate the generality of the proposed Hamiltonian through examples of systems with various total angular momenta relevant to ultracold-gas physics.
In particular, in Sec.~\ref{sec:SpecificEqualMomenta}, we reproduce the known results for atoms with equal total angular momenta in the bosonic and fermionic cases, while in Sec.~\ref{sec:SpecificMixture} we study interatomic interaction in mixtures of atoms with different total angular momenta for both statistics, yielding new results.
Finally, Sec. \ref{sec:Conclusion} summarizes the main results and conclusions of the work.

\section{Interaction Hamiltonian of a homonuclear atomic mixture}\label{sec:GeneralHamiltonian}

\subsection{Physical aspects of interactions between atoms in different states}\label{ssec:Aspects}

To describe interactions in a many-body system representing a homonuclear mixture of atoms in different quantum states, it is necessary to first specify the state of an individual atom. 
This can be done by choosing an appropriate basis determined by the external conditions. 
In experiments with ultracold gases, the regime of weak magnetic fields is typically realized. In this case, the natural basis is characterized by the following quantum numbers: the total atomic angular momentum $f_i$, its projection $m_i$, the orbital angular momentum $l_i$, spin $s_i$ of the electronic shell, and nuclear spin. Therefore, the total atomic angular momentum $f_i$ is composed of nuclear spin and the total angular momentum of the electron shell. 
Each such set of the above quantum numbers uniquely specifies the distribution of electrons in the atomic shell and, at the same time, determines the electric and magnetic properties of the atom. 

These properties, in turn, can be determined through measurements of the atomic multipole moments such as the monopole, dipole, quadrupole moments, etc., which arise as the coefficients in the large-distance expansion of the electromagnetic potential \cite{Landau_FT}. 
Thus, the interaction between two sufficiently distant atoms may be represented as the interaction between their multipole moments. 
Such an approximation is quite appropriate for dilute systems such as ultracold gases. 
However, within the framework of quantum mechanics the situation becomes more complicated. 
Interactions are commonly divided into two types: direct and exchange. 
The direct interaction corresponds to ordinary scattering, in which only the energies of the colliding atoms are redistributed. 
The exchange interaction, by contrast, has a purely quantum nature and consists in the “exchange” of certain quantum numbers between identical particles. 
In quantum electrodynamics, this process may be interpreted as the exchange of virtual photon(s) \cite{Craig_MolecularQE_1984}. 
Since a photon carries an integer number of units of angular momentum, the angular momentum of the atoms in such a process remains either integer (bosons) or half-integer (fermions); in other words, the quantum  statistics of the atoms is preserved.

In addition, within a quantum-mechanical description, the interaction can be represented as a weighted sum of contributions from scattering channels, each characterized by the total angular momentum $|f_{1}-f_{2}|\leq F \leq f_1+f_2$ of the two atoms. 
The equivalence between the descriptions in terms of multipolar exchange and scattering-channel contributions can be established using the Wigner-Eckart theorem, which relates the multipole components of the atomic fields to the components of irreducible spherical tensor operators.

\subsection{Benefits of the Spherical Tensor Operator Approach}
\label{ssec:STO}

Before constructing the Hamiltonian, we briefly discuss several useful properties of irreducible spherical tensor operators. 
The operators $T^{K}_{\kappa}$, where $K$ and $\kappa$ denote the rank and the component, respectively, act in the space of internal atomic states and form a complete orthogonal basis. 
The matrix elements of $T^{K}_{\kappa}$ are determined by the Clebsch–Gordan coefficients $C^{f_{1}m_{1}}_{f_{2}m_{2}\,K\kappa}$ \cite{Varshalovich_1988}:
\begin{gather}
    \left<f_{1}m_{1}|T^{K}_{\kappa}|f_{2}m_{2}\right>=(T^{K}_{\kappa})_{m_{1}m_{2}}=(-1)^{2K}C^{f_{1}m_{1}}_{f_{2}m_{2}\,K\kappa}\frac{\left<f_{1}||T^{K}||f_{2}\right>}{\sqrt{2f_{1}+1}}
    \nonumber
    \\
    =
    (-1)^{f_{1}-m_{1}}
    \left(\begin{matrix}f_{1} & K & f_{2} \\ -m_{1} & \kappa & m_{2} \end{matrix}\right)
    \left<f_{1}||T^{K}||f_{2}\right>,
    \label{eq:Wig-Eck}
\end{gather}
where the double-bar matrix element, independent of the magnetic quantum numbers $m_{1}$, $m_{2}$, and $\kappa$, is known as the reduced matrix element and the $2\times 3$ array represents a 3$j$ symbol. 
Equation \eqref{eq:Wig-Eck} expresses the essence of the celebrated Wigner–Eckart theorem. The standard Condon-Shortley phase convention establishes the following property for the reduced matrix elements \cite{Edmonds_1996}: 
\begin{equation}
    \left<f_{1}||T^{K}||f_{2}\right>
    =
    (-1)^{f_1-f_2}
    \left<f_{2}||T^{K}||f_{1}\right>. 
    \label{eq:RedME}
\end{equation}
We use the Racah normalization, which together with Eq.~\eqref{eq:RedME} yield
\begin{equation}
    \left<f_{1}||T^{K}||f_{2}\right>
    =
    \left\{
    \begin{array}{cc}
         \sqrt{2K+1}, &  f_1\leq f_2\\
         (-1)^{f_1-f_2}\sqrt{2K+1}, & f_1>f_2.
    \end{array}
    \right. 
    \label{eq:RedME2}
\end{equation}

Spherical tensor operators enable the interaction Hamiltonian to be decomposed into a sum of all distinct multipolar exchange terms: scalar ($K=0$), dipole ($K=1$), quadrupole ($K=2$), and higher-order moments. 
Each term contributes in a specific way to the behavior of the system. 
In particular, for the interaction of two atoms, the Hamiltonian can be expressed as a scalar product of tensor operators:
\begin{equation}
    H^{\textrm{int}}
    = 
    \sum_{K} U^K\sum_{\kappa=-K}^{K} (-1)^{\kappa} \, T^K_{\kappa}(1) \, T^K_{-\kappa}(2)
    ,
    \label{eq:GeneralHamiltonianThroughSTO}
\end{equation}
where $U^{K}$ is the interaction potential  corresponding to the exchange of multipole moments of rank $K$ and labels $(1)$ and $(2)$ refer to the first and second atom, respectively. 
The representation \eqref{eq:GeneralHamiltonianThroughSTO} ensures that the Hamiltonian is a scalar under SO(3) and thus conserves angular momentum.
The contraction of spherical tensor operators in Eq.~\eqref{eq:GeneralHamiltonianThroughSTO} assumes an isotropic interaction, which is dominated by low-energy $s$-wave scattering (the relative orbital angular momentum $L=0$). This is a natural assumption for ultracold gases. The anisotropic contributions associated with higher partial waves (e.g., $p$-wave scattering) would require coupling the spherical tensor operators to spatial degrees of freedom via spherical harmonics.

Such representation of Hamiltonian significantly simplifies the analysis of the interaction matrix elements, as it allows the direct application of the formal tools of angular-momentum theory \cite{Wigner_1959,Racah_1959,Brink_1968,Edmonds_1996,Sakurai_2020}, including the Clebsch–Gordan rules, Racah coefficients, and the Wigner–Eckart theorem.
The tensor-operator formalism is particularly useful for many-body systems, where interactions can be projected onto channels with definite total angular momentum.
It also enables the efficient construction of the interaction Hamiltonian with isotropic, as in the present case, or anisotropic contributions \cite{Krems_JPhysChemA_2004,Krems_InRevPhysChem_2005}.

The rank and component of a tensor operator uniquely determine the structure of its matrix representation, which is difficult to achieve using SU(N) generators without additional conventions. 
Moreover, the close connection with the Wigner–Eckart theorem allows the construction of the tensor-operator representation to be generalized to arbitrary angular momentum. 
For interactions between atoms with different $f_{i}$, the corresponding representations, as shown below, may take the form of rectangular matrices, which cannot be obtained within the SU(N) generator framework.
\subsection{Construction of interaction Hamiltonian}
\label{subsec:InterHam}

We now proceed to the construction of the many-body Hamiltonian for a system of interacting homonuclear ultracold atoms in different internal quantum states $i=\{f_{i},\lambda_{i}\}$, where $\lambda_{i}=\{l_i,s_i,\dots\}$
denotes the set of the remaining atomic quantum numbers. The Hamiltonian of such a system can be written as
\begin{equation}
    H
    =
    H_0
    +
    \sum_{j\geq i}
    H^{\textrm{int}}_{ij}
    ,
\end{equation}
where $H_0$ includes kinetic energy and the coupling to an external field, while $H^{\textrm{int}}_{ij}$ describes all interactions between atoms, both within a single component and between different components. The interaction Hamiltonian between atoms in the same state, $H^{\textrm{int}}_{ii}$, is obtained by setting the corresponding quantum numbers equal.

Let us consider a pair of atoms in different quantum states $1=\{f_1,\lambda_{1}\}$ and $2=\{f_2,\lambda_{2}\}$, taking into account their indistinguishability. To this end, we symmetrize the wave function of the pair with respect to the simultaneous permutation of spatial coordinates and spin variables:
\begin{equation}
    \Psi^{FM}_{12}
    (\textbf{x},\textbf{x}')
    =
    \frac{1}{\sqrt{2(1+\delta_{12})}}
    \sum_{m_{1},m_{2}}
    \left[
    C^{FM}_{f_1m_{1}\,f_2m_{2}}
    \psi_{m_1}(\mathbf{x})
    \psi_{m_2}(\mathbf{x}')
    +
    C^{FM}_{f_2m_{2}\,f_1m_{1}}
    \psi_{m_2}(\mathbf{x})
    \psi_{m_1}(\mathbf{x}')
    \right].
   \label{eq:waveFuncF}
\end{equation}
Evidently, it has the following parity property:
\begin{equation}
    \Psi^{FM}_{12}
    (\textbf{x},\textbf{x}')
    =(-1)^{f_{1}+f_{2}-F}\Psi^{FM}_{21}
    (\textbf{x}',\textbf{x}).
    \label{eq:waveFuncParity}
\end{equation}
Here $|f_1-f_2|\leq F\leq f_1+f_2$ is the total angular momentum, $-F\leq M\leq F$ is its projection, and $C^{FM}_{f_1m_{1}\,f_2m_{2}}$ are the Clebsch–Gordan coefficients. The wave function in Eq.~\eqref{eq:waveFuncF} is normalized provided that the single-particle states are orthonormal, 
\begin{gather}
\int d \textbf{x} \psi_{1m_{1}}^{*}(\textbf{x})\psi_{2m_{2}}(\textbf{x})=\delta_{12}\delta_{m_1m_2}, \quad 
\psi_{m_{i}}(\textbf{x})\equiv\psi_{im_{i}}(\textbf{x}).
\end{gather}

We can now write the corresponding interaction Hamiltonian for atoms in quantum states 1 and 2:
\begin{gather}
    H^{\textrm{int}}_{12}
    =
    \frac{1}{2}
   \int 
    d\mathbf{x}d\mathbf{x}'
    \sum_{F,M}
    U^{F}_{12}(\mathbf{x}-\mathbf{x}')
    \Psi^{FM*}_{12}(\mathbf{x},\mathbf{x}')
    \Psi^{FM}_{12}(\mathbf{x},\mathbf{x}') 
    \nonumber\\
    =
    \frac{1}{2(1+\delta_{12})}
    \int d\mathbf{x}d\mathbf{x'}
    \sum_{m_{1},m_{2},m_{1}',m_{2}'}
    \left[
        \psi^{*}_{m_1}(\textbf{x})
        \psi^{*}_{m_2}(\textbf{x}')
        \Phi^{\textrm{Proj}}_{m_1m_2m_1'm_2'}(\mathbf{x}-\mathbf{x}')
        \psi_{m_2'}(\textbf{x}')
        \psi_{m_1'}(\textbf{x})
    \right.
    \nonumber\\
    \left.
        +
        \psi^{*}_{m_1}(\textbf{x})
        \psi^{*}_{m_2}(\textbf{x}')
        \Phi^{\textrm{Mom}}_{m_1m_2m_2'm_1'}(\mathbf{x}-\mathbf{x}')
        \psi_{m_1'}(\textbf{x}')
        \psi_{m_2'}(\textbf{x})
    \right]
    ,
    \label{eq:HamiltonianThroughCGC}
\end{gather}
where
\begin{gather}
    \Phi^{{\textrm{Proj}}}_{m_1m_2m_1'm_2'}(\mathbf{x}-\mathbf{x}')
    =
    \sum_{F,M}
    U^{F}_{12}(\mathbf{x}-\mathbf{x}')
    C^{FM}_{f_1m_{1}\,f_2m_{2}}C^{FM}_{f_1m_{1}'\,f_2m_{2}'}
    ,
    \label{eq:InterF1}
    \\
    \Phi^{{\textrm{Mom}}}_{m_1m_2m_2'm_1'}(\mathbf{x}-\mathbf{x}')
    =
    \sum_{F,M}
    U^{F}_{12}(\mathbf{x}-\mathbf{x}')
    C^{FM}_{f_1m_{1}\,f_2m_{2}}
    C^{FM}_{f_2m_{2}'\,f_1m_{1}'}
    \label{eq:InterF2}
\end{gather}
and $U^{F}_{12}(\mathbf{x}-\mathbf{x}')$ is the interaction potential of two atoms in the channel characterized by $F$.
As expected, the resulting Hamiltonian includes both direct and all possible exchange interactions between atoms in states 1 and 2. In particular, the interaction function $\Phi^{{\textrm{Proj}}}_{m_1m_2m_1'm_2'}$ describes the exchange of angular-momentum projections, whereas $\Phi^{{\textrm{Mom}}}_{m_1m_2m_2'm_1'}$ accounts for the exchange of the total angular momenta themselves.

It should be noted that two-body interactions are also characterized by the relative orbital angular momentum, $L$. 
As we already mentioned, in ultracold atomic gases, which are the subject of the present study, collisions occur at low energies, so the relative motion of the atoms involves zero orbital angular momentum ($L=0$), corresponding to $s$-wave scattering. 
This imposes restrictions on the allowed scattering channels characterized by $F$. 
According to the symmetry property of the two-particle wave function given by Eq.~\eqref{eq:waveFuncParity}, the quantity $f_{1}+f_{2}-F$ must be even for bosons and odd for fermions.

Next, noting that, according to the Condon-Shortley convention, the Clebsch-Gordan coefficients are real,
$
    C^{FM}_{f_1m_{1},f_2m_{2}}=\left<1,2|FM\right>=\left<FM|1,2\right>,
$
the interaction functions given by Eqs.~\eqref{eq:InterF1} and \eqref{eq:InterF2}
can be expressed in terms of the projection operator $P^{F}=|FM\rangle\langle FM|$ onto the two-particle state with total angular momentum $F$:
\begin{gather}
    \Phi^{\textrm{Proj}}_{m_1m_2m_1'm_2'}(\mathbf{x}-\mathbf{x}')
    =
    \sum_{F}
    U^{F}_{12}(\mathbf{x}-\mathbf{x}')
    P^{F}_{m_1m_2m_1'm_2'}
    \label{eq:ProjOperProj}
    ,\\
    \Phi^{\textrm{Mom}}_{m_1m_2m_2'm_1'}(\mathbf{x}-\mathbf{x}')
    =
    \sum_{F}
    U^{F}_{12}(\mathbf{x}-\mathbf{x}')
    P^{F}_{m_1m_2m_2'm_1'}
    .
     % \Phi_{122'1'}(\mathbf{x}-\mathbf{x}')
    % =
    % \sum_{F}
    % (-1)^{F}
    % U^{F}_{\Lambda f_1f_2}(\mathbf{x}-\mathbf{x}')
    % \left<
    %     1,2
    % \right|
    %     P^{F}
    % \left|
    %     1',2'
    % \right>
    % .
    \label{eq:ProjOperMom}
\end{gather}

As noted in subsection \ref{ssec:STO}, it is more convenient and physically transparent to describe the interaction in terms of spherical tensor operators. 
Accordingly, the Hamiltonian \eqref{eq:HamiltonianThroughCGC} can be rewritten in a form similar to that of \eqref{eq:GeneralHamiltonianThroughSTO}. 
This can be achieved by using the Wigner–Eckart theorem (see \eqref{eq:Wig-Eck}), which establishes the connection between the matrix elements of spherical tensor operators and the Clebsch–Gordan coefficients. 
However, before its direct application, it is necessary to properly “repack” the indices in the Clebsch–Gordan coefficients, making use of their symmetry properties, so that the matrix elements of the tensor operators correctly reproduce the structure of the interaction Hamiltonian, including both the exchange of projections and the total angular momenta themselves.

First, it follows from Eq. \eqref{eq:InterF1} that the quantity $\Phi^{\textrm{Proj}}_{m_1 m_2 m_1' m_2'}$ corresponds to a physical process in which the atomic total angular momentum $f_i$ remains unchanged, while the projections $m_i$ may vary. To reconcile the form of expression \eqref{eq:InterF1} with its physical interpretation, we perform several mathematical transformations. In particular, using the relations \eqref{eq:Sum_KG} and \eqref{eq:KG_Prop} together with the Wigner–Eckart theorem \eqref{eq:Wig-Eck}, we obtain
\begin{align}
    &\Phi^{\textrm{Proj}}_{m_1m_2m_1'm_2'}(\mathbf{x}-\mathbf{x}')
    =
    \nonumber\\
    &\sum_{F}
    U^{F}_{12}(\mathbf{x}-\mathbf{x}')
    (2F+1)
    (-1)^{f_1+ f_2+F}
    \sum_{K,\kappa}
         \begin{Bmatrix} 
            f_1 & f_1 & K \\ 
            f_2 & f_2 & F 
         \end{Bmatrix}
    (-1)^{\kappa}
    (T^{K}_{\kappa})_{m_1m_1'}
    (T^{K}_{-\kappa})_{m_2m_2'}. 
    \label{eq:PhiExchProj}
\end{align}
Here the reduced matrix element is taken in accordance to Eq.~\eqref{eq:RedME} and \eqref{eq:RedME2}, and $2\times 3$ array specifies the Wigner 6$j$ symbol. Spherical tensor operators $T^{K}_{\kappa}$ act in the spaces with angular momenta $f_{1}$ and $f_{2}$. Since the same rank $K$ is used for operators in both spaces, it must satisfy simultaneously the conditions $0\leq K\leq 2f_{1}$ and $0\leq K\leq 2f_{2}$, yielding the allowed range $0\leq K\leq 2\min(f_{1},f_{2})$ with $-K\leq\kappa\leq K$. 

An analogous procedure based on Eqs. \eqref{eq:KG_Prop1}–\eqref{eq:Sum_KG} must also be applied to the expression \eqref{eq:InterF2}, which, according to Eq.~\eqref{eq:HamiltonianThroughCGC}, physically corresponds to the exchange of total angular momenta between the atoms. 
Thus, we arrive at the following expression:
\begin{align}
    &\Phi^{\textrm{Mom}}_{m_1m_2m_2'm_1'}(\mathbf{x}-\mathbf{x}')
    =
    \nonumber\\
    &\sum_{F}
    U^{F}_{12}(\mathbf{x}-\mathbf{x}')
    (2F+1)
    (-1)^{f_1+f_2-F}
    \sum_{K,\kappa}
         \begin{Bmatrix} 
         f_1 & f_2 & K \\ f_1 & f_2 & F 
         \end{Bmatrix}
    (-1)^{\kappa}
    (T^{K}_{\kappa})_{m_1m_2'}
    (T^{K}_{-\kappa})_{m_2m_1'}
    ,
    \label{eq:PhiExchMom}
\end{align}
In contrast to Eq.~\eqref{eq:PhiExchProj}, the tensor operators in Eq. \eqref{eq:PhiExchMom} have rank $|f_1-f_2|\leq K\leq f_1+f_2$ and components $-K\leq\kappa\leq K$.

In the second-quantization formalism, the single-particle wave functions in the Hamiltonian \eqref{eq:HamiltonianThroughCGC} are replaced by field operators:
\begin{gather*}
    \psi_{m_i}(\textbf{x})\to
    \hat{\psi}_{m_i}(\textbf{x})
    =
    \frac{1}{\sqrt{V}}
    \sum_{\textbf{p}}
    e^{i\textbf{px}/\hbar}
    \hat{a}_{\textbf{p}m_i}
    ,
    \\
    \psi_{m_i}^{*}(\textbf{x})\to
    \hat{\psi}^{\dagger}_{m_i}(\textbf{x})
    =
    \frac{1}{\sqrt{V}}
    \sum_{\textbf{p}}
    e^{-i\textbf{px}/\hbar}
    \hat{a}^{\dagger}_{\textbf{p}m_i}
    ,
\end{gather*}
where $V$ denotes the system volume. As a result, the corresponding interaction operator can be written as
\begin{gather}
    H^{\textrm{int}}_{12}
    =
    \frac{1}{2(1+\delta_{12})V}
    \sum_{\textbf{p}_{1}...\textbf{p}_{4}}
    \sum_{m_1,m_2,m_1',m_2'}
    \bigg[
    \tilde{\Phi}^{\textrm{Proj}}_{m_1m_2m_1'm_2'}(\mathbf{p}_1-\mathbf{p}_4)
    a^{\dagger}_{\textbf{p}_1m_1}
    a^{\dagger}_{\textbf{p}_2m_2}
    a_{\textbf{p}_3m_2'}
    a_{\textbf{p}_4m_1'}
    \nonumber
    \\
+  \tilde{\Phi}^{\textrm{Mom}}_{m_1m_2m_2'm_1'}(\mathbf{p}_1-\mathbf{p}_3)
    a^{\dagger}_{\textbf{p}_1m_1}
    a^{\dagger}_{\textbf{p}_2m_2}
    a_{\textbf{p}_4m_1'}
    a_{\textbf{p}_3m_2'}
    \bigg]
   \delta_{\textbf{p}_1+\textbf{p}_2,\textbf{p}_3+\textbf{p}_4}
   ,
   \label{eq:InterHamFourier}
\end{gather}
where the following Fourier transforms are defined:
\begin{gather}
    \tilde{\Phi}^{\textrm{Proj}}_{m_1m_2m_1'm_2'}(\textbf{p})=\int d{\textbf x}
    \Phi^{\textrm{Proj}}_{m_1m_2m_1'm_2'}(\textbf{x})e^{-i\textbf{px}/\hbar},
    \nonumber
    \\
    \tilde{\Phi}^{\textrm{Mom}}_{m_1m_2m_2'm_1'}(\textbf{p})=\int d{\textbf x}
    \Phi^{\textrm{Mom}}_{m_1m_2m_2'm_1'}(\textbf{x})e^{-i\textbf{px}/\hbar}
    \label{eq:PhiForier}
\end{gather}
and the interaction functions $\Phi^{\textrm{Proj}}_{m_1m_2m_1'm_2'}(\textbf{x})$ and $\Phi^{\textrm{Mom}}_{m_1m_2m_2'm_1'}(\textbf{x})$ are given by Eqs.~\eqref{eq:PhiExchProj} and \eqref{eq:PhiExchMom}, respectively. 
The creation $a^{\dagger}_{\textbf{p}_im_i}$ and annihilation $a_{\textbf{p}_im_i}$  operators obey the standard commutation or anticommutation relations, depending on the particle statistics.

\begin{figure}[htb]
    \centering
    \includegraphics[width=164mm]{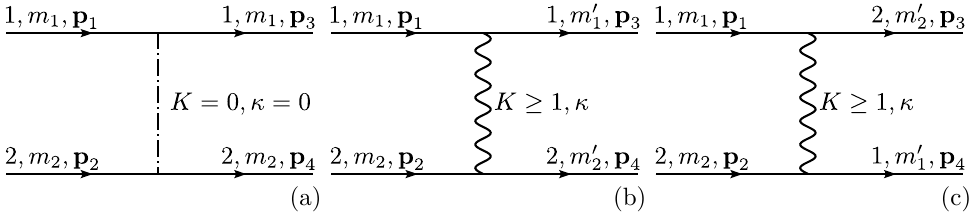}
    \caption{
    Feynman diagrams of the interaction terms in the Hamiltonian, represented as an effective exchange of: (a) a virtual scalar particle, (b) a virtual photon without total angular momentum transfer, and (c) a virtual photon with total angular momentum transfer.
    }
    \label{fig:feynmanDiagrams}
\end{figure}

The obtained expression for the interaction Hamiltonian in terms of tensor operators provides a transparent physical interpretation of the underlying interaction processes. These processes are mediated by the exchange of a virtual particle and can be illustrated by the diagrams shown in Fig.~\ref{fig:feynmanDiagrams}. 
The term with $K=0$ corresponds to the direct interaction and may be associated with the exchange of a virtual scalar particle (rather than a photon).
For $K\geq 1$, the tensor operators {\it effectively} describe the exchange of a virtual photon between the interacting atoms, which is a natural concept within quantum electrodynamics \cite{Craig_MolecularQE_1984}. Therefore, the rank $K$ of a tensor operator determines the total angular momentum carried by the exchanged virtual particle, while its component $\kappa$ specifies the corresponding projection. 
Moreover, the sets of quantum numbers $\lambda_{1}=\{l_1,s_1,\dots\}$ and $\lambda_{2}=\{l_2,s_2,\dots\}$, which specify the interaction potential $U_{12}^{F}(\textbf{x}-\textbf{x}')$, allow one to establish the spatial parity of the interaction process as $(-1)^{l_{1}-l_{2}}$. This parity factor, in turn, provides a direct criterion for identifying the nature of the exchanged multipole moments: electric or magnetic. Finally, by comparing Eqs.~\eqref{eq:PhiExchProj} and \eqref{eq:PhiExchMom} with Eq.~\eqref{eq:GeneralHamiltonianThroughSTO}, one can extract the interaction potential $U^{K}$. Evidently, it differs for each virtual-particle exchange process.

As noted above, this Hamiltonian can be applied to a broad class of problems in the physics of quantum many-body systems. It describes the interaction between two atoms with arbitrary total angular momenta $f_{1}$ and $f_{2}$, either integer or half-integer. Several important particular realizations of this Hamiltonian in the context of quantum gases are discussed in the next section.

\section{Gas of atoms in equal internal state, \texorpdfstring{$f_{1}=f_{2}$ and $\lambda_{1}=\lambda_{2}$}{f1=f2, la1=la2}}
\label{sec:SpecificEqualMomenta}
In this section we demonstrate that the general interaction functions derived above reproduce the well-known results for atoms with equal total angular momenta. 
Then comparing Eqs. \eqref{eq:PhiExchProj} and \eqref{eq:PhiExchMom}, we find that the two functions have identical forms up to a permutation of the last two indices, $m_{1}'\leftrightarrow m_{2}'$. 
Therefore, it is  sufficient to determine the explicit form of only one of them for the fermionic and bosonic cases considered below. For definiteness, we choose the function $ \Phi^{\textrm{Proj}}_{m_1m_2m_1'm_2'}(\mathbf{x}-\mathbf{x}')$.

It is also convenient to introduce the spherical components of the total angular momentum operator,
\begin{equation}
    f_{\pm}
    =
    \mp
    \frac{1}{\sqrt{2}}(f_{x}\pm if_{y}), \quad f_{0}=f_{z},
    \label{eq:SphericaComp_f}
\end{equation}
where $f_{x}$, $f_{y}$ and $f_{z}$ are its Cartesian components. The relations in Eq.~\eqref{eq:SphericaComp_f} are general and hold for any value of the total angular momentum \cite{Varshalovich_1988,Sakurai_2020}. 

In this and subsequent sections, distinct matrix indices ($v_i, n_i, m_i, w_i$) are introduced, where the number of values for each index equals the multiplicity of the corresponding total atomic angular momentum.

\subsection{Fermionic gas of atoms with \texorpdfstring{$f_{1}=f_{2}=1/2$}{f1=f2=1/2}}

Now we proceed to the particular case of identical fermionic atoms with total angular momentum $f_{1}=f_{2}=1/2$ ($F,\,K=0,1$). To this end, we note that for the matrix equivalents of tensor operators defined by the Wigner-Eckart theorem (see Eq.~\eqref{eq:Wig-Eck}), the following relations hold: 
\begin{equation*}
    T^{0}_{0}=\frac{1}{\sqrt{2}}I, \quad 
    T^{1}_{0}=\sqrt{2}f_{0}, 
    \quad 
    T^{1}_{1}=\sqrt{2}f_{+},
    \quad
    T^{1}_{-1}=\sqrt{2}f_{-},
\end{equation*}
where $I$ is the identity operator. Evaluating the corresponding $6j$ symbols and performing straightforward algebraic manipulations, we obtain from Eq.~\eqref{eq:PhiExchProj}
\begin{gather}
     \Phi_{v_{1}v_{2}v_{1}'v_{2}'}^{\textrm{Proj}}(\textbf{x}-\textbf{x}')
     =
     \left(
     \frac{1}{4}
     U^{0}_{1/2,1/2}(\textbf{x}-\textbf{x}')
     +
     \frac{3}{4}
     U^{1}_{1/2,1/2}(\textbf{x}-\textbf{x}')
     \right)
     \delta_{v_{1}v_{1}'}
     \delta_{v_{2}v_{2}'}
     \nonumber\\
     -
     \left(
     \vphantom{\frac12}
     U^{0}_{1/2,1/2}(\textbf{x}-\textbf{x}')-U^{1}_{1/2,1/2}(\textbf{x}-\textbf{x}')
     \right)
     \textbf{f}_{v_{1}v_{1}'}
     \textbf{f}_{v_{2}v_{2}'},
     \label{eq:Phi_1/2}
\end{gather} 
where indices $v_{i},v_{i}'$ run over $2f_{i}+1=2$ values.
Expression \eqref{eq:Phi_1/2} is in agreement with the well-known interaction Hamiltonian for identical spin-1/2 particles (see, e.g., \cite{Akhiezer_SpinWaves_1968}). 

In the case of $s$-wave scattering, the $F=1$ channel is forbidden. Therefore, imposing the condition $P^{1}=0$ on the corresponding projection operator, Eq.~\eqref{eq:ProjOperProj}) yields
\begin{gather*}
     \Phi_{v_{1}v_{2}v_{1}'v_{2}'}^{\textrm{Proj}}(\textbf{x}-\textbf{x}')
     =
     U^{0}_{1/2,1/2}(\textbf{x}-\textbf{x}')
     \delta_{v_{1}v_{1}'}
     \delta_{v_{2}v_{2}'}.
\end{gather*}
Thus, upon substituting the obtained result into Eq.~\eqref{eq:InterHamFourier}, 
one finds that only atoms with opposite spin projections participate in the scattering process.

\subsection{Bosonic gas of atoms with \texorpdfstring{$f_{1}=f_{2}=1$}{f1=f2=1}}

A second important case in the physics of ultracold gases is that of interacting identical bosons with total angular momentum $f_{1}=f_{2}=1$ ($F,\,K=0,1,2$). The corresponding interaction function is obtained directly  from Eq.~\eqref{eq:PhiExchProj}:
% \begin{flalign}
%     \Phi_{n_{1}n_{2}n_{1}'n_{2}'}^{\textrm{Proj}}(\textbf{x}-\textbf{x}')
%     =
%     \left(
%         \frac{1}{3}U^{0}_{11}(\textbf{x}-\textbf{x}')+U^{1}_{11}(\textbf{x}-\textbf{x}')+\frac{5}{3}U^{2}_{11}(\textbf{x}-\textbf{x}')
%     \right)
%     (T^{0}_{0})_{n_{1}n_{1}'}
%     (T^{0}_{0})_{n_{2}n_{2}'}
%     \nonumber\\
%     +
%     \left(
%         -\frac{1}{3}U^{0}_{11}(\textbf{x}-\textbf{x}')
%         -
%         \frac{1}{2}U^{1}_{11}(\textbf{x}-\textbf{x}')+\frac{5}{6}U^{2}_{11}(\textbf{x}-\textbf{x}')
%     \right)
%     \sum_{\kappa}(-1)^{\kappa}(T^{1}_{\kappa})_{n_{1}n_{1}'}(T^{1}_{-\kappa})_{n_{2}n_{2}'}
%     \nonumber\\
%     +
%     \left(
%         \frac{1}{3}U^{0}_{11}(\textbf{x}-\textbf{x}')
%         -
%         \frac{1}{2}U^{1}_{11}(\textbf{x}-\textbf{x}')
%         +
%         \frac{1}{6}U^{2}_{11}(\textbf{x}-\textbf{x}')
%     \right)
%     \sum_{\kappa}(-1)^{\kappa}(T^{2}_{\kappa})_{n_{1}n_{1}'}(T^{2}_{-\kappa})_{n_{2}n_{2}'}.
%     \label{eq:Phi11}
% \end{flalign}
\begin{equation}
    \Phi_{n_{1}n_{2}n_{1}'n_{2}'}^{\textrm{Proj}}(\textbf{x}-\textbf{x}')
    =
    \sum_{F,K,\kappa}
    (-1)^{K+\kappa}
    U^{F}_{1,1}(\textbf{x}-\textbf{x}')
    A^{FK}_{1,1}
    (T^{K}_{\kappa})_{n_{1}n_{1}'}
    (T^{K}_{-\kappa})_{n_{2}n_{2}'},
    \label{eq:Phi11}
\end{equation}
where $n_{i}, n_{i}'$ run through $2f_{i}+1=3$ values and $A_{1,1}^{FK}$ is the matrix composed of the coefficients derived from the evaluation of the relevant $6j$ symbols,
\begin{equation*}
    A^{FK}_{1,1}
    =
    \left(
        \begin{array}{rrr}
            \dfrac13 & \dfrac13 & \dfrac13 \\[1em]
            1 & \dfrac12 & -\dfrac12 \\[1em]
            \dfrac53 & -\dfrac56 & \dfrac16 \\
        \end{array}
    \right)
    .
\end{equation*}
Equation \eqref{eq:Phi11} contains a scalar product of rank-two tensor operators. Physically, the spherical tensor operator $T^{2}_{\kappa}$ is equivalent to the traceless quadrupole tensor $Q^{ik}=f_{i}f_{k}+f_{k}f_{i}-4/3\delta_{ik}$, since their components are related by a linear transformation. Furthermore, the explicit expansion of the invariant $Q^{ik}(1)Q^{ki}(2)$ yields a linear combination of three independent scalar operators --- the identity, the bilinear invariant ${\textbf f}(1)\cdot{\textbf f}(2)$, and the biquadratic invariant $({\textbf f}(1)\cdot{\textbf f}(2))^{2}$. Therefore, the Hamiltonian with the interaction function given by Eq.~\eqref{eq:Phi11} is equivalent to the well-known form featuring bilinear and biquadratic exchange (see, e.g., \cite{Bulakhov_JPhysA_2022}).

From a comparison of Eqs.~\eqref{eq:ProjOperProj} and \eqref{eq:Phi11}, we can also find the projection operators onto the two-particle states with total angular momentum $F=0,1,2$: 
\begin{equation*}
    P^{F}_{n_{1}n_{2}n_{1}'n_{2}'}
    =
    \sum_{K,\kappa}
    (-1)^{K+\kappa}
    A^{FK}_{1,1}
    (T^{K}_{\kappa})_{n_{1}n_{1}'}
    (T^{K}_{-\kappa})_{n_{2}n_{2}'}.
\end{equation*}
However, since we are interested in $s$-wave scattering of two identical atoms, the channel with $F=1$ is forbidden by the symmetry requirement of the wave function \eqref{eq:waveFuncF} under exchange of atoms (see also \cite{Ueda_PhysRep_2012}). This leads to the condition  $P^{1}=0$, which allows one to eliminate the contributions of rank-2 tensor operators in Eq.~\eqref{eq:Phi11}. As a result, we find
\begin{align}
    \Phi_{n_1n_2n_1'n_2'}^{\textrm{Proj}}(\textbf{x}-\textbf{x}')
    =
    &
    \left(
        \frac{1}{3}
        U^{0}_{1,1}(\textbf{x}-\textbf{x}')
        +
        \frac{2}{3}U^{2}_{1,1}(\textbf{x}-\textbf{x}')
    \right)
    \delta_{n_{1}{n_{1}'}}
    \delta_{n_{2}{n_{2}'}}
    \nonumber
    \\
    +
    &
    \left(
        \frac{1}{3}
        U^{2}_{1,1}(\textbf{x}-\textbf{x}')
        -
        \frac{1}{3}
        U^{0}_{1,1}(\textbf{x}-\textbf{x}')
    \right)
    \textbf{f}_{n_{1}n_{1}'}
    \textbf{f}_{n_{2}n_{2}'}.
\label{eq:Phi_f_1}
\end{align}
% \begin{gather}
%     \Phi_{m_1m_2m_1'm_2'}^{\textrm{Proj}}(\textbf{x}-\textbf{x}')
%     =
%     \delta(\textbf{x}-\textbf{x}')
% \left[
%     \frac{U^{0}+2U^{2}}{3}
%     \delta_{m_{1}{m_{1}'}}
%     \delta_{m_{2}{m_{2}'}}
%     +
%     \frac{U^{2}-U^{0}}{3}
%     \textbf{f}_{m_{1}m_{1}'}
%     \textbf{f}_{m_{2}m_{2}'}
% \right].
% \label{eq:Phi_f_1}
% \end{gather}
In deriving Eq.~\eqref{eq:Phi_f_1}, we used the fact that the components of the tensor operators are related to the identity operator $I$ and the spherical components of the spin operator $f_{\kappa}$ by 
\begin{equation*}
    T^{0}_{0}
    =
    \frac{1}{\sqrt{3}}I
    ,\quad 
    T^{1}_{0}
    =
    \frac{1}{\sqrt{2}}f_{0}
    ,\quad 
    T^{1}_{1}
    =
    \frac{1}{\sqrt{2}}f_{+}
    ,\quad
    T^{1}_{-1}
    =
    \frac{1}{\sqrt{2}}f_{-} 
    .
    \label{eq:Spin1SpherComp}
\end{equation*}
The interaction function \eqref{eq:Phi_f_1} gives rise to a Hamiltonian widely used in studies of magnetic states of weakly interacting atomic gases with a Bose-Einstein condensate \cite{Ohmi_JPSJ_1998,Ho_PRL_1998,Ueda_PhysRep_2012}.

\section{Mixture of atoms in different internal states \texorpdfstring{$f_{1}\neq f_{2}$}{f1=f2}} 
\label{sec:SpecificMixture}
Finally, we consider the most nontrivial case of mixtures of interacting atoms in different internal states. 
For example, the states may have the same total angular momentum but different values of $\lambda_{i}$. 
However, the most illustrative applications of the general expressions presented in Subsec.~\ref{subsec:InterHam} arise in the case of different total angular momenta, since the quantities $\lambda_{i}$ do not explicitly enter the numerical coefficients. 

To illustrate the rectangular matrix representation, we study examples of bosonic ${}^{4}$He and fermionic ${}^{3}$He atoms in their ground and metastable states \cite{Hodgman_PRL_2009}. 
Strontium \cite{Yasuda_PhysRevLett_2004}, ytterbium  \cite{Ishiyama_PhysRevLett_2023} and recently magneto-optically trapped titanium atoms \cite{Eustice_PhysRev_2025} are also excellent candidates for applying our formalism due to the long lifetimes of their metastable states: $\approx 520$ s, $\approx 2$ s and $\approx 28\cdot 10^3$ s, respectively. 
However, we omit them here because their large total angular momenta would complicate the expressions without providing new physical insight.

\subsection{Bosonic mixture of atoms with \texorpdfstring{$f_{1}=0,\ f_{2}=1$}{f1=0, f2=1}}

Here we consider a bosonic mixture of atoms with $f_{1}=0$ and $f_{2}=1$. This choice of angular momenta is not only convenient for illustrative purposes but is also of potential practical interest. As noted in the introduction, these values of the total angular momenta correspond to the ground and excited states of the ${}^{4}$He atom. 
As is well known, the excited state with $f_{2}=1$ is the longest-lived metastable atomic state, with a lifetime of approximately $7870\, \mathrm{s}$ (more than two hours) \cite{Hodgman_PRL_2009}.   

We begin with the interaction between atoms within a single component (i.e., with identical $f_i$). For atoms with $f_{1}=0$, we obtain 
\begin{gather*}
    \Phi_{m_{1}m_{2}m_{1}'m_{2}'}^{\textrm{Proj}}(\textbf{x}-\textbf{x}')
    =
    U^{0}_{00}(\textbf{x}-\textbf{x}')(T^{0}_{0})_{m_{1}m_{1}'}(T^{0}_{0})_{m_{2}m_{2}'},
    \quad
     (T^{0}_{0})_{m_i,m_i'}
    =
    1,
\end{gather*}
where $m_{i}, m_{i}'$ run over $2f_{1}+1=1$ value only.
The interaction function for the component with $f_{2}=1$ is determined above by Eqs.~\eqref{eq:Phi11} and \eqref{eq:Phi_f_1}.

In addition, the system under consideration exhibits interactions between components with different total angular momenta, $f_{1}=0$ and $f_{2}=1$. According to Eq.~\eqref{eq:PhiExchProj}, the interaction function for the exchange of projections takes the form ($F,K=1$): 
\begin{equation}
    \Phi_{m_1n_2m_1'n_2'}^{\textrm{Proj}}(\textbf{x}-\textbf{x}')
    =
    \sqrt{3}
    U_{01}^{1}(\textbf{x}-\textbf{x}')
    (T^{0}_{0})_{m_1m_1'}
    (T^{0}_{0})_{n_2n_2'}, 
    \label{eq:Phi01ExchProj}
\end{equation}
while that for the exchange of angular momentum is given by Eq.~\eqref{eq:PhiExchMom} ($F=1,K=0$):
\begin{equation}
    \Phi_{m_1n_2n_2'm_1'}^{\textrm{Mom}}(\textbf{x}-\textbf{x}')
    =
    U_{01}^{1}(\textbf{x}-\textbf{x}')
    \sum_{\kappa}
    (-1)^{\kappa}
    (T^{1}_{\kappa})_{m_1n_2'}
    (T^{1}_{-\kappa})_{n_2m_1'},
    \label{eq:Phi01ExchMom}
\end{equation}
where $n_{i}, n_{i}'$ take $2f_{2}+1=3$ values. As we see, the expressions obtained above from Eqs.~\eqref{eq:PhiExchProj} and \eqref{eq:PhiExchMom} contribute differently to the many-body Hamiltonian (a mere interchange of $m_1'$ and $n_2'$ is not sufficient), in contrast to the case of equal $f_{i}$. For completeness and in view of potential applications of these results, we also provide the explicit form of the matrix equivalents of the spherical tensor operators appearing in Eq.~\eqref{eq:Phi01ExchMom},   
\begin{gather*}
   (T^{1}_{-1})_{m_{1}n_{2}'}
    =
    \begin{pmatrix}
        1 & 0 & 0
    \end{pmatrix}, 
\quad
    (T^{1}_{0})_{m_{1}n_{2}'}
    =
    \begin{pmatrix}
        0 & -1 & 0
    \end{pmatrix},
\quad
    (T^{1}_{1})_{m_{1}n_{2}'}
    =
    \begin{pmatrix}
        0 & 0 & 1
    \end{pmatrix},
\nonumber\\
      (T^{1}_{-1})_{n_{2}m_{1}'}
    =
    \begin{pmatrix}
        0 \\ 0 \\ -1
    \end{pmatrix}, 
\quad
   ( T^{1}_{0})_{n_{2}m_{1}'}
    =
    \begin{pmatrix}
        0 \\ -1 \\ 0
    \end{pmatrix}
    ,
\quad
    (T^{1}_{1})_{n_{2}m_{1}'}
    =
    \begin{pmatrix}
        -1 \\ 0 \\ 0
    \end{pmatrix}
.
\end{gather*}

\subsection{Fermionic mixture of atoms with \texorpdfstring{$f_{1}=1/2,\ f_{2}=3/2$}{f1=1/2, f2=3/2}}

As another physically relevant example, we consider the fermionic isotope ${}^{3}$He. In contrast to the previously studied ${}^{4}$He atoms, the nonzero nuclear spin gives rise to the hyperfine ground and excited sublevels with $f_{1}=1/2$ and $f_{2}=3/2$, respectively. Since the lifetime of the metastable state is governed primarily by the electronic structure, the $f_{2}=3/2$ hyperfine state of ${}^{3}$He is also long-lived, with a lifetime comparable to that of ${}^{4}$He. 

The interaction between atoms within the $f_{1}=1/2$ component has already been determined by Eq.~\eqref{eq:Phi_f_1}, while that within the $f_{2}=3/2$ component is given by Eq.~\eqref{eq:PhiExchProj}, which in the present case yields
\begin{equation}
    \Phi_{w_{1}w_{2}w_{1}'w_{2}'}^{\textrm{Proj}}(\textbf{x}-\textbf{x}')
    =
    \sum_{F,K,\kappa}
    (-1)^{K+\kappa}
    U^{F}_{3/2,3/2}(\textbf{x}-\textbf{x}')
    A^{FK}_{3/2,3/2}
    (T^{K}_{\kappa})_{w_{1}w_{1}'}
    (T^{K}_{-\kappa})_{w_{2}w_{2}'},
    \label{eq:PhiExchProj3/2}
\end{equation}
where 
\begin{equation*}
    A^{FK}_{3/2,3/2}
    =
    \left(\begin{array}{rrrr}
        \dfrac14 & \dfrac14 & \dfrac14 & \dfrac14 \\[2ex]
        \dfrac34 & \dfrac{11}{20} & \dfrac{3}{20} & -\dfrac{9}{20} \\[2ex]
        \dfrac54 & \dfrac14 & -\dfrac34 & \dfrac14 \\[2ex]
        \dfrac74 & -\dfrac{21}{20} & \dfrac{7}{20} & -\dfrac{1}{20} \\
    \end{array}\right)
    .
\end{equation*}
and $w_{i}, \ w_{i}'$ take $2f_2+1=4$ values. The matrix equivalents of the tensor operators $(T^{K}_{\kappa})_{w_{i}w_{i}'}$ and the matrix $A^{FK}_{3/2,3/2}$ can be found in Ref.~\cite{Bulakhov_JPhysA_2023}.
The respective projection operators onto the two-particle states with total angular momentum $F=0,1,2,3$ are given by
\begin{equation*}
P_{w_{1}w_{2}w_{1}'w_{2}'}^{F}
=
    \sum_{K,\kappa}
    (-1)^{K+\kappa}
    A^{FK}_{3/2,3/2}
    (T^{K}_{\kappa})_{w_{1}w_{1}'}
    (T^{K}_{-\kappa})_{w_{2}w_{2}'}.
\end{equation*}
In case of low-energy collisions characterized by $s$-wave scattering, two channels with $F=1,3$ are already closed, so that $P^{1}=0$ and $P^{3}=0$. As above, these conditions allow us to eliminate the contributions of the rank-two and rank-three tensor operators in general Eq.~\eqref{eq:PhiExchProj3/2}. As a result, we obtain
\begin{align}
    \Phi_{w_{1}w_{2}w_{1}'w_{2}'}^{\textrm{Proj}}(\textbf{x}-\textbf{x}')
    =&
    \left(
        \frac{5}{4}
        U^{2}_{3/2,3/2}(\textbf{x}-\textbf{x}')
        -
        \frac{1}{4}
        U^{0}_{3/2,3/2}(\textbf{x}-\textbf{x}')
    \right)
    \delta_{w_{1}w_{1}'}\delta_{w_{2}w_{2}'}
    \nonumber\\
    +&
     \left(
        \frac{1}{3}
        U^{2}_{3/2,3/2}(\textbf{x}-\textbf{x}')
        -
        \frac{1}{3}
        U^{0}_{3/2,3/2}(\textbf{x}-\textbf{x}')
    \right)
    \textbf{f}_{w_{1}w_{1}'}\textbf{f}_{w_{2}w_{2}'}
    .
\end{align}
Therefore, in the case of $s$-wave scattering, the interatomic interaction between atoms with total angular momentum $f_{2}=3/2$ can be completely described in terms of spin operators alone, with no contribution from higher-rank multipole operators. However, higher-rank multipole operators contribute when $f_{i}\geq 2$.

In contrast to the intracomponent case, the intercomponent interaction is characterized by two distinct functions defined by Eqs.~\eqref{eq:PhiExchProj} and \eqref{eq:PhiExchMom}. Evaluating the corresponding $6j$ symbols, Eq.~\eqref{eq:PhiExchProj} leads to the interaction function describing the exchange of projections, 
\begin{multline}
    \Phi_{v_{1}w_{2}v_{1}'w_{2}'}^{\textrm{Proj}}(\textbf{x}-\textbf{x}')
    =
    \left(
        \frac{3}{2\sqrt{2}}
        U^{1}_{1/2,3/2}(\textbf{x}-\textbf{x}')
        +
        \frac{5}{2\sqrt{2}}
        U^{2}_{1/2,3/2}(\textbf{x}-\textbf{x}')
    \right)
    (T^{0}_{0})_{v_{1}v_{1}'}
    (T^{0}_{0})_{w_{2}w_{2}'}
    % \nonumber
    \\
    +
    \left(
        -\frac{5}{2\sqrt{10}}
        U^{1}_{1/2,3/2}(\textbf{x}-\textbf{x}')
        +
        \frac{5}{2\sqrt{10}}
        U^{2}_{1/2,3/2}(\textbf{x}-\textbf{x}')
    \right)
    \sum_{\kappa}
    (-1)^{\kappa}
    (T^{1}_{\kappa})_{v_{1}v_{1}'}
    (T^{1}_{-\kappa})_{w_{2}w_{2}'}
    ,
    \label{eq:PhiProj25}
\end{multline}
where $v_{i},\ v_{i}'$ run over $2f_1+1=2$ values. In accordance with Eq.~\eqref{eq:ProjOperProj}, the corresponding projection operator for the closed scattering channel $F=2$ (due to $s$-wave scattering) has the form
\begin{flalign*}
    % P^{1}_{v_1w_2v_1'w_2'}
    % =
    % \sum_{\kappa}
    % (-1)^{\kappa}
    % \left[
    %     \frac{1}{2\sqrt{2}}
    %     (T^{0}_{\kappa})_{v_1v_1'}
    %     (T^{0}_{-\kappa})_{w_2w_2'}
    %     -
    %     \frac{5}{6\sqrt{10}}(T^{1}_{\kappa})_{v_1v_1'}
    %     (T^{1}_{-\kappa})_{w_2w_2'}
    % \right]
    % ,
    % \\
    P^{2}_{v_1w_2v_1'w_2'}
    =
    \sum_{\kappa}
    (-1)^{\kappa}
    \left[
        \frac{5}{2\sqrt{2}}
        (T^{0}_{\kappa})_{v_1v_1'}
        (T^{0}_{-\kappa})_{w_2w_2'}
        +
        \frac{5}{2\sqrt{10}}
        (T^{1}_{\kappa})_{v_1v_1'}
        (T^{1}_{-\kappa})_{w_2w_2'}
    \right]
    .
\end{flalign*}
Therefore, eliminating in Eq.~\eqref{eq:PhiProj25} the rank-0 operators by using $P^{2}=0$, we obtain
\begin{equation}
    \Phi_{v_{1}w_{2}v_{1}'w_{2}'}^{\textrm{Proj}}(\textbf{x}-\textbf{x}')
    =
    -\frac{4}{\sqrt{10}}
    U^{1}_{1/2,3/2}(\textbf{x}-\textbf{x}')
    \sum_{\kappa}
    (-1)^{\kappa}
    (T^{1}_{\kappa})_{v_{1}v_{1}'}
    (T^{1}_{-\kappa})_{w_{2}w_{2}'}
    .
    \label{eq:PhiProj25_Fin}
\end{equation}

In a similar manner, following Eqs.~\eqref{eq:PhiExchMom} and \eqref{eq:ProjOperMom}, we have
\begin{multline}
    \Phi_{v_{1}w_{2}w_{2}'v_{1}'}^{\textrm{Mom}}(\textbf{x}-\textbf{x}')
    =
    \left(
        \frac{1}{4}
        U^{1}_{1/2,3/2}(\textbf{x}-\textbf{x}')
        +
        \frac{5}{4}
        U^{2}_{1/2,3/2}(\textbf{x}-\textbf{x}')
    \right)
    \sum_{\kappa}
    (-1)^{\kappa}
    (T^{1}_{\kappa})_{v_{1}w_{2}'}
    (T^{1}_{-\kappa})_{w_{2}v_{1}'}
    % \nonumber
    \\
    +
    \left(
        -
        \frac{3}{4}
        U^{1}_{1/2,3/2}(\textbf{x}-\textbf{x}')
        +
        \frac{1}{4}
        U^{2}_{1/2,3/2}(\textbf{x}-\textbf{x}')
    \right)
    \sum_{\kappa}
    (-1)^{\kappa}
    (T^{2}_{\kappa})_{v_{1}w_{2}'}
    (T^{2}_{-\kappa})_{w_{2}v_{1}'}.
    \label{eq:PhiMom25}
\end{multline}
with
\begin{equation*}
    P^{2}_{v_1w_2w_2'v_1'}
    =
    \sum_{\kappa}
    (-1)^{\kappa}
    \left[
        \frac{5}{4}
        (T^{1}_{\kappa})_{v_1w_2'}
        (T^{1}_{-\kappa})_{w_2v_1'}
        +
        \frac{1}{4}
        (T^{2}_{\kappa})_{v_1w_2'}
        (T^{2}_{-\kappa})_{w_2v_1'}
    \right]
    .
\end{equation*}
% \begin{flalign*}
%     % P^{1}_{v_1w_2w_2'v_1'}
%     % =
%     % \sum_{\kappa}
%     % (-1)^{\kappa}
%     % \left[
%     %     -
%     %     \frac{1}{12}
%     %     (T^{1}_{\kappa})_{v_1w_2'}
%     %     (T^{1}_{-\kappa})_{w_2v_1'}
%     %     +
%     %     \frac{1}{4}
%     %     (T^{2}_{\kappa})_{v_1w_2'}
%     %     (T^{2}_{-\kappa})_{w_2v_1'}
%     % \right]
%     % ,
%     % \\
%     P^{2}_{v_1w_2w_2'v_1'}
%     =
%     \sum_{\kappa}
%     (-1)^{\kappa}
%     \left[
%         -
%         \frac{5}{4}
%         (T^{1}_{\kappa})_{v_1w_2'}
%         (T^{1}_{-\kappa})_{w_2v_1'}
%         -
%         \frac{1}{4}
%         (T^{2}_{\kappa})_{v_1w_2'}
%         (T^{2}_{-\kappa})_{w_2v_1'}
%     \right]
%     .
% \end{flalign*}
Again, employing the condition $P^{2}=0$ to eliminate rank-two operators gives
\begin{equation}
    \Phi_{v_{1}w_{2}w_{2}'v_{1}'}^{\textrm{Mom}}(\textbf{x}-\textbf{x}')
    =
    4
    U^{1}_{1/2,3/2}(\textbf{x}-\textbf{x}')
    \sum_{\kappa}
    (-1)^{\kappa}
    (T^{1}_{\kappa})_{v_{1}w_{2}'}
    (T^{1}_{-\kappa})_{w_{2}v_{1}'}
    .
    \label{eq:PhiMom25_Fin}
\end{equation}
Similarly to the previously studied bosonic atoms with $f_{1}\neq f_{2}$, the exchange of total angular momentum (see Eq.~\eqref{eq:PhiMom25_Fin}) is described by tensor operators represented by the following rectangular matrices:
\begin{align*}
    &(T^{1}_{0})_{v_1w_2'}
    =
    \begin{pmatrix}
     0 & -\frac{1}{\sqrt{2}} & 0 & 0 \\
     0 & 0 & -\frac{1}{\sqrt{2}} & 0 \\
    \end{pmatrix}
    ,
    &&(T^{1}_{0})_{w_2v_1'}
    =
    \begin{pmatrix}
     0 & 0 \\
     -\frac{1}{\sqrt{2}} & 0 \\
     0 & -\frac{1}{\sqrt{2}} \\
     0 & 0 \\
    \end{pmatrix}
    ,
    \\
    &(T^{1}_{-1})_{v_1w_2'}
    =
    \begin{pmatrix}
     \frac{\sqrt{3}}{2} & 0 & 0 & 0 \\
     0 & \frac{1}{2} & 0 & 0 \\
    \end{pmatrix}
    ,
    &&(T^{1}_{-1})_{w_2v_1'}=
    \begin{pmatrix}
     0 & 0 \\
     0 & 0 \\
     -\frac{1}{2} & 0 \\
     0 & -\frac{\sqrt{3}}{2} \\
    \end{pmatrix}
    ,\\
    &(T^{1}_{1})_{v_1w_2'}
    =
    \begin{pmatrix}
     0 & 0 & \frac{1}{2} & 0 \\
     0 & 0 & 0 & \frac{\sqrt{3}}{2} \\
    \end{pmatrix}
    ,
    &&(T^{1}_{1})_{w_2v_1'}=
    \begin{pmatrix}
     -\frac{\sqrt{3}}{2} & 0 \\
     0 & -\frac{1}{2} \\
     0 & 0 \\
     0 & 0 \\
    \end{pmatrix}
    .
\end{align*}
We note that the choice of operator rank to be eliminated in \eqref{eq:PhiProj25} and \eqref{eq:PhiMom25} is arbitrary and affects only the explicit form of the resulting expressions. 

\section{Conclusion}\label{sec:Conclusion}

We have proposed a general method for constructing the many-body Hamiltonian of pairwise interactions in homonuclear mixtures of atoms occupying states with different values of the total angular momentum and other quantum numbers. 
The approach is based on the formalism of irreducible spherical tensor operators, which provides the Hamiltonian with a transparent physical structure, naturally incorporates all scattering channels, and allows one to include all multipolar exchange interactions. 
The latter describe both the exchange of angular-momentum projections and the exchange of the angular momenta themselves. 
We have illustrated the generality of the proposed Hamiltonian through examples of systems relevant to ultracold atomic gases, composed of interacting atoms with both equal and unequal total angular momenta.

It is worth noting that alternative mathematical frameworks for describing multipolar exchange also exist, including the use of Stevens operators \cite{Stevens_1952,Rudowicz_2004}, representations in terms of SU(N) generators \cite{Peletminskii_PLA_2020,Bulakhov_JPhysA_2021}, and other approaches \cite{BernatskaJPhysA_2009,Kiselev_EPL_2013}.
However, these formalisms do not possess all the advantages of spherical tensor operators, such as universality, a transparent classification in terms of rank and component, and the direct use of angular-momentum algebra. In particular, by universality we refer to the simplicity of constructing rectangular representations (see, e.g., Sec.~\ref{sec:SpecificMixture}).

Moreover, the obtained expressions \eqref{eq:PhiExchProj} and  \eqref{eq:PhiExchMom} for the interaction functions written in terms of tensor operators admit a clear physical interpretation. 
The exchange interaction can be viewed as an effective exchange mediated by a virtual particle (see Fig. \ref{fig:feynmanDiagrams}). 
The rank $K$ of the tensor operator determines the total angular momentum of the exchanged particle, while its component $\kappa$ specifies the corresponding projection. The transfer of angular momentum or its projection by a virtual photon is, in general, a natural concept within quantum electrodynamics \cite{Craig_MolecularQE_1984}.

Our description assumes that the atomic state is characterized by the total angular momentum (i.e., the coupled basis). 
This approximation is valid for external magnetic fields smaller than the hyperfine splitting energies, as well as for low densities where the intrinsic atomic fields and close collisions can be neglected.
The opposite case, where the description is formulated in the uncoupled basis, also arises in studies of Feshbach resonances and anisotropic interactions \cite{Krems_book_2009,Krems_InRevPhysChem_2005,Krems_JPhysChemA_2004}. 
In general, these two approaches complement each other, as they describe interactions at different spatial scales and can, in principle, be incorporated into the Hamiltonian simultaneously.  

In summary, the resulting Hamiltonian provides a convenient and universal framework for theoretical studies of a wide range of quantum many-body phenomena in bosonic and fermionic high-spin atomic gases, where multipolar exchange induced by angular momentum plays an essential role. 
%It may also serve as a starting point for constructing models of interacting multicomponent ultracold atomic systems, including lattice models.
%\warning{add }
It may also serve as a starting point for constructing models of interacting multicomponent ultracold atomic systems. 
Specifically, the continuum interaction potentials in our Hamiltonian can be directly extracted from the asymptotic boundary conditions provided by Multichannel Quantum Defect Theory (MQDT) \cite{Grimm_RevModPhys_2010,Laskowski_PhysRevA_2023}. Furthermore, applying the tight-binding approximation by integrating this Hamiltonian with Wannier functions readily yields generalized Hubbard models for multicomponent gases in optical lattices \cite{Zoller_PRL_1998,Gorshkov_NP_2010}.
However, such applications could divert attention from the main features of the proposed approach and, therefore, go beyond the scope of the present work.

\section*{Acknowledgment}
Authors acknowledge support from the National Research Foundation of Ukraine under the initiative “Excellent Fundamental Science in Ukraine 2027-2029” through the project "Development of theoretical methods for thorough analysis of collective properties of quantum many-body systems" (No.~2026.06/0012). 
% A.S. Peletminskii acknowledges support from the Simons Foundation and Simons Foundation International through the Centers of Excellence in Ukraine grant (No.~PD-COE-00032387).

\begin{appendix}
\section{Properties of the Clebsch-Gordan coefficients}
Following \cite{Varshalovich_1988}, we summarize several properties of the Clebsch–Gordan coefficients. In doing so, we adopt the convention of using Latin letters to denote total angular momenta and Greek letters to denote their projections. The Clebsch–Gordan coefficients then satisfy the following symmetry relations:
\begin{gather}
C^{c\gamma}_{a\alpha \, b \beta}=(-1)^{a
    +b-c}\, C^{c\gamma}_{b \beta \, a\alpha}, \label{eq:KG_Prop1}
    \\ 
 C^{c\gamma}_{a\alpha \, b \beta}=(-1)^{b+\beta}\sqrt{\frac{2c+1}{2a+1}}C^{a\alpha}_{b-\beta\, c\gamma}
    =(-1)^{a-\alpha}\sqrt{\frac{2c+1}{2b+1}}C^{b-\beta}_{a\alpha \, c-\gamma}=(-1)^{b+\beta}\sqrt{\frac{2c+1}{2a+1}}C^{a-\alpha}_{c-\gamma \, b\beta}.
    \label{eq:KG_Prop}
\end{gather}
The sum involving Clebsch–Gordan coefficients and a 6$j$ symbol is given by 
\begin{equation}
    \sum_{c\gamma}(-1)^{c+d-\beta-\varphi}(2d+1)\,C^{c\gamma}_{a\alpha \, b \beta}C^{c\gamma}_{f-\varphi \, e \epsilon}\begin{Bmatrix} a & b & c \\ e & f & d \end{Bmatrix}=\sum_{\delta} C^{d\delta}_{a\alpha \, f \varphi}C^{d\delta}_{b-\beta \, e \epsilon}.
    \label{eq:Sum_KG}
\end{equation}
This formula coincides with that given in Ref. \cite{Varshalovich_1988} once the triangle rule is taken into account. 

% The Clebsch–Gordan coefficients determine the matrix elements of the irreducible spherical tensor operator $T^{K}_{\kappa}$:
% \begin{gather}
%     \left<a\alpha|T^{K}_{\kappa}|b\beta\right>=(T^{K}_{\kappa})_{\alpha\beta}=(-1)^{2K}C^{a\alpha}_{b\beta\,K\kappa}\frac{\left<a||T^{K}||b\right>}{\sqrt{2a+1}}
%     \\
%     =
%     (-1)^{a-\alpha}
%     \left(\begin{matrix} a & K & b \\ -\alpha & \kappa & \beta \end{matrix}\right)
%     \left<a||T^{K}||b\right>,
%     \label{eq:Wig-Eck}
% \end{gather}
% where the double-bar matrix element, independent of magnetic numbers $\alpha$, $\beta$, and $\kappa$, is known as the reduced matrix element and $2\times 3$ matrix represents 3$j$ symbol. Relation \eqref{eq:Wig-Eck} captures the essence of the celebrated Wigner–Eckart theorem.

\end{appendix}

\bibliographystyle{apsrev}
\bibliography{main}

\end{document}